# Astrophysics in the Laboratory — The CBM Experiment at FAIR

**Peter Senger** [1, 2, *]

[1]  Facility for Antiproton and Ion Research, 64291 Darmstadt, Germany; p.senger@gsi.de

[2]  National Research Nuclear University MEPhI, 115409 Moscow, Russia

* for the CBM collaboration

**Abstract:** The future "Facility for Antiproton and Ion Research" (FAIR) is an accelerator-based international center for fundamental and applied research, which presently is under construction in Darmstadt, Germany. An important part of the program is devoted to questions related to astrophysics, including the origin of elements in the universe and the properties of strongly interacting matter under extreme conditions, which are relevant for our understanding of the structure of neutron stars and the dynamics of supernova explosions and neutron star mergers. The Compressed Baryonic Matter (CBM) experiment at FAIR is designed to measure promising observables in high-energy heavy-ion collisions, which are expected to be sensitive to the high-density equation-of-state (EOS) of nuclear matter and to new phases of QCD matter at high densities. The CBM physics program, the relevant observables and the experimental setup will be discussed.

**Keywords:** heavy-ion collisions; nuclear matter equation-of-state; QCD phase diagram

## 1. The Future Facility for Antiproton and Ion Research (FAIR)

The Facility for Antiproton and Ion Research (FAIR) will provide unique research opportunities in hadron and nuclear physics, in atomic physics and nuclear astrophysics, in materials research, plasma physics and radiation biophysics, including developments of novel medical treatments and applications for space science [1]. The layout of the planned facility is illustrated in Figure 1. The FAIR start version comprises the synchrotron SIS100, which will provide high-intensity beams of protons up to an energy of 29 GeV and nuclei up to 15 A GeV for $Z/A = 0.5$. Gold or Uranium beams will be available with kinetic energies up to 11 A GeV. High-intensity secondary beams will be produced by a large acceptance Superconducting Fragment Separator, which collects very efficiently rare isotopes created in reactions with the primary beams. The properties of these isotopes will be investigated by the experimental facilities of the NUSTAR collaboration (**Nu**clear **St**ructure, **A**strophysics and **R**eactions), in order to shed light on the various paths of nucleosynthesis in the universe. The High-Energy Storage Ring (HESR) will cool and accelerate intense secondary beams of antiprotons, which will be used for hadron physics experiments by the PANDA collaboration (Anti**P**roton **An**nihilation at **Da**rmstadt). A dedicated cave will host experimental facilities for experiments in atomic, plasma and biophysics and in material science. The detector system of the Compressed Baryonic Matter (CBM) experiment is designed to investigate high-energy heavy-ion collisions, with the goal to explore the properties of high-density QCD matter. The construction of the accelerator components and civil construction is progressing. According to the actual planning, installation and commissioning of the experiments is planned during 2022–2024 and in 2025 FAIR will be deliver first beams.

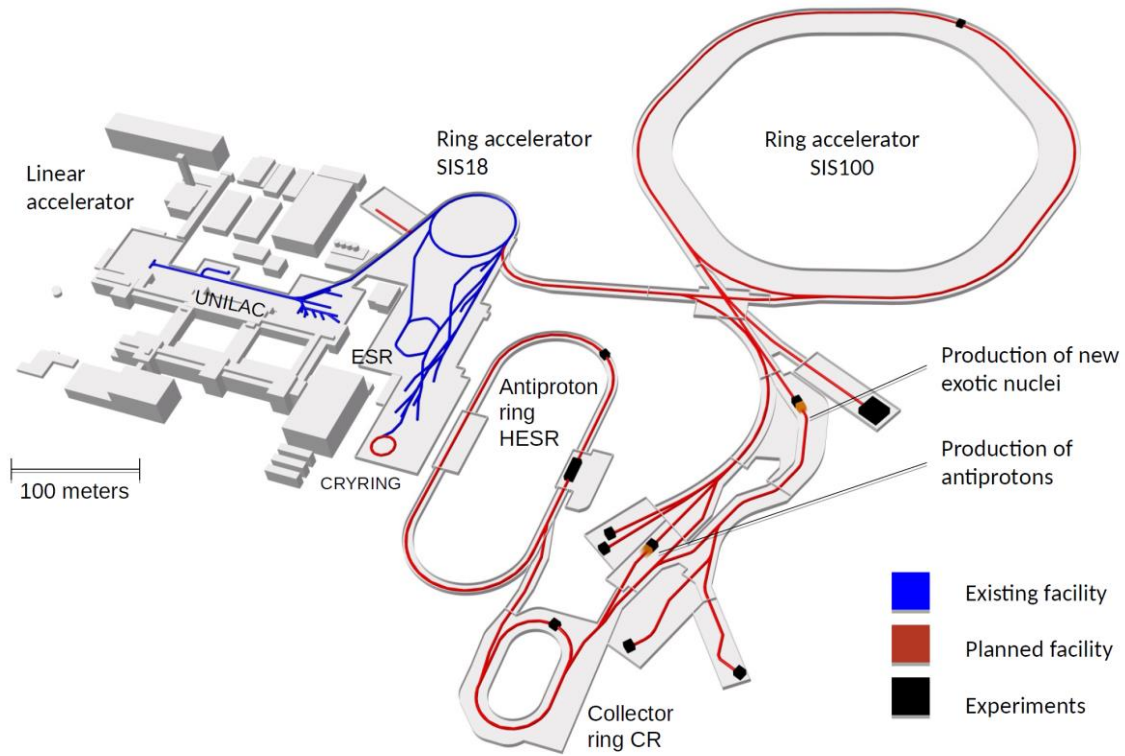

**Figure 1.** Layout of the future Facility for Antiproton and Ion Research (FAIR) [1].

## 2. Exploring the Origin of Elements

The research program conducted by the NUSTAR collaboration is substantially related to the investigation of the origin of elements in the universe. The nuclei involved in the astrophysical processes are often short-lived isotopes with significant neutron deficiency or excess. These nuclei have to be artificially produced in the laboratory, which requires the development of novel techniques and instrumentation. FAIR will provide a unique combination of high-energy primary beams from the SIS100 and the Super-Fragment Separator followed by storage rings equipped with state-of-the-art instrumentation These experimental setups will allow the identification of pure secondary beams of short-lived nuclei, often for the first time, pushing the frontier of knowledge deeper into the yet unexplored regions of the nuclear chart.

Figure 2 illustrates the various paths of nucleosynthesis within the nuclear chart and sketches the corresponding astrophysical processes and sites. The nuclei up to the mass of iron are created by fusion processes in stars. In a supernova explosion, these nuclei are expelled into space becoming the birth material of future star generations. The nuclei with masses close to and beyond the proton dripline are synthesized in binary systems of a sun and a neutron star by the rapid proton capture (rp) and proton capture (p) processes. The slow neutron (s) process proceeds in very heavy stars along the valley of stability and produces about 50% of all heavy elements up to Bismuth. About half of the elements heavier than iron and all trans-actinides in the Universe are produced during the astrophysical r-process by a sequence of rapid neutron captures and beta decays. The process occurs in an environment with extreme neutron densities, as they are encountered in merging neutron stars and core-collapse supernovae, where the latter will likely only contribute to the synthesis of the lighter r-process nuclei up to the mass range A = 130. The nuclei on the r-process path are all short-lived and have such large neutron excess that most of them have not yet been produced in the laboratory. Hence their properties have to be modelled introducing significant uncertainties into r-process simulations. This situation will drastically improve once FAIR will be operational.

A breakthrough in our understanding of the r-process was triggered by a merger of two neutron stars, which has been observed by the LIGO and Virgo collaborations via the measurement of the gravitational waves [2,3]. The detection of the gravitational wave signal was followed by a series of multi-messenger observations, including the electro-magnetic radiation. This "kilonova" signal has been predicted in 2010 as a consequence of the radioactive decay of heavy elements synthesized in ejecta from the merger [4]. In conclusion, neutron star mergers probably are the major source of heavy elements including gold, platinum and uranium.

The most important nuclear properties required for r-process simulations are neutron separation energies (masses) which, for given environment conditions, determine the nuclei on the r-process path and beta half-lives which influence the dynamics of the process. The Super-FRS will be the most powerful in-flight separator to measure both masses and half-lives of rare isotopes of all elements up to uranium, which can be produced and spatially separated within some hundred nanoseconds. Therefore, thus very short-lived nuclei can be studied efficiently.

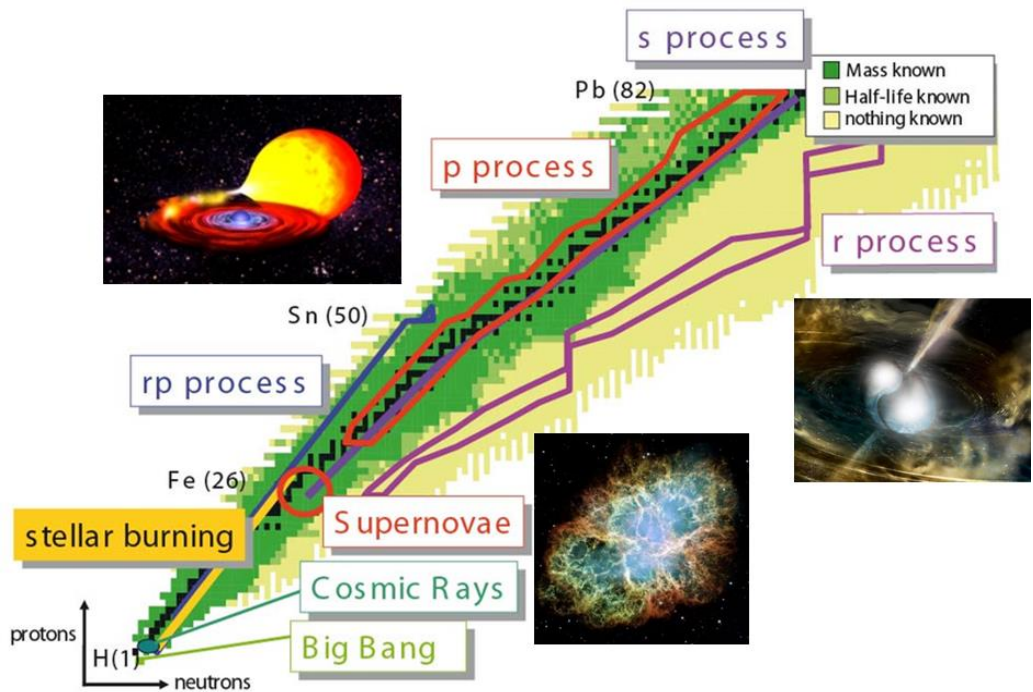

**Figure 2.** Sketch of the nuclear chart including the various paths of nucleosynthesis. The rapid proton capture (rp) process and the proton capture (p) process happen in binary systems consisting of a sun and a neutron star. The rapid neutron capture (r) process takes place in supernova explosions and neutron star mergers. The subfigures illustrate the corresponding astrophysical processes and sites.

**3. Exploring the Properties and Phases of High-Density QCD Matter.**

Heavy-ion collisions at beam energies as available at FAIR are very well suited to create dense baryonic matter, although only for an extremely short time and of tiny dimensions. In the reaction zone of two heavy nuclei colliding with kinetic beam energies between 2 and 11 A GeV, the baryon density exceeds 2–6 times saturation density [5]. In nature, such densities are expected to exist in the core of massive neutron stars, in supernovae and in neutron star mergers. This is illustrated in Figure 3, which sketches the phase diagram of QCD matter as function of temperature and net-baryon density [6]. The conditions like in the early universe shortly after the big bang, where matter and antimatter exists in equal amounts, are realized at extremely high temperatures and at vanishing net-baryon density. Here, a smooth cross-over from hadronic to quark matter is predicted by the fundamental theory of strong interaction, Quantum Chromo Dynamics (QCD), at a pseudo-critical temperature of about 155 MeV [7,8]. This value coincides with the freeze-out temperature of the fireball, when fitting a statistical model to the yields of particles produced in heavy-ion collisions at ultra-relativistic beam energies, as available at the Relativistic Heavy Ion Collider (RHIC) at BNL in USA and at the Large Hadron Collider (LHC) at CERN in Switzerland. Unfortunately, QCD theory is not applicable at high net-baryon densities and the only predictions about possible structures in the QCD phase diagram, like the location of a 1$^{st}$ order phase transition with a critical endpoint or new exotic phases of QCD matter, are provided by models [9,10]. Figure 3 sketches the conjectured landmarks such as a 1$^{st}$ order phase transition and the critical endpoint and indicates the densities and temperatures expected in the core of neutron stars and in the collision zone of neutron star mergers. A review of the theoretical concepts and the experimental approaches devoted to the exploration of the QCD phase diagram in the region of high net-baryon densities is given in the CBM Physics Book [11].

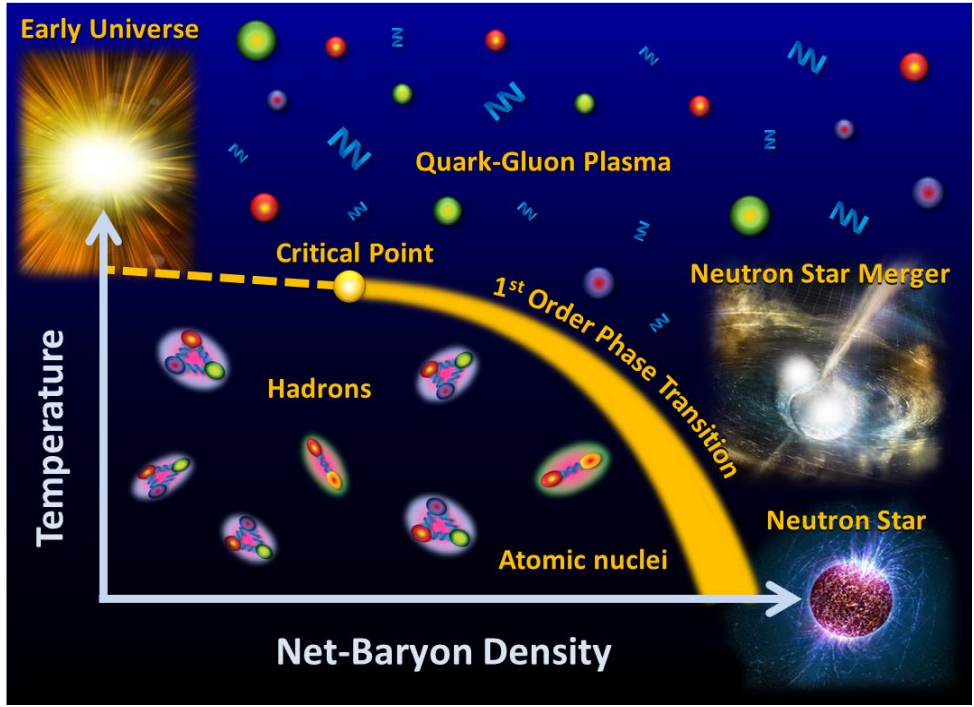

**Figure 3.** Sketch of the phase diagram of nuclear matter as function of temperature and net-baryon density [6]. The yellow dashed line and the yellow region indicate the locations of a crossover and a first-order phase transition, respectively, separating the regions of hadronic matter and matter consisting of quarks and gluons. Very high temperatures at zero net-baryon densities did exist in the early universe. Matter at high net-baryon densities albeit at different temperatures is realized in the core of neutron stars and in neutron star mergers.

This article reviews promising future experimental investigations of heavy-ion collision, which are expected to shed light on the high-density nuclear matter equation-of-state, possibly reveal new phases of QCD matter at high net-baryon densities and contribute to the solution of the hyperon puzzle in neutron stars by studying (double-) lambda hyper-nuclei.

*3.1. The High-Density Nuclear-Matter Equation-of-State*

The high-density equation-of-state (EOS) is a fundamental property of nuclear matter, which governs the structure of neutron stars and the dynamics of neutron star mergers and core collapse supernovae. Therefore, various astrophysics experiments try to constrain the EOS at high densities. For example, the Neutron Star Interior Composition Explorer (NICER) detector located at the International Space Station measures time-resolved x-ray spectra emitted from hot spots of neutron stars, in order to extract simultaneously information on their mass and radius [12]. Moreover, constraints on the radii of neutron stars have been derived by multi-messenger observations of neutron star mergers in combination with nuclear theory [13]. Once the mass and radius of a nucleon star has been measured with high accuracy, the high-density EOS can be calculated via the Tolman-Oppenheimer-Volkoff

(TOF) equation. The complementary approach is to measure diagnostic probes sensitive to the EOS in high-energy heavy-ion experiments and compare to results of transport calculations, which take into account in-medium effects and relativistic nucleon-nucleon potentials.

The EOS describes the relation between density, pressure, volume, temperature, energy and isospin asymmetry. The pressure can be written as the energy per nucleon

$$E_A(\rho,\delta) = E_A(\rho,0) + E_{sym}(\rho)\cdot\delta^2 + O(\delta^4), \tag{1}$$

with the asymmetry parameter $\delta = (\rho_n - \rho_p)/\rho$. Symmetric matter is stable around saturation density $\rho_0$ with a binding energy of $E/A(\rho_o) = -16$ MeV, the slope $\delta(E/A)(\rho_o)/\delta\rho = 0$ and the curvature $K_{nm} = 9\rho^2 \delta^2(E/A)/\delta\rho^2$ with $K_{nm}$ the nuclear incompressibility. At saturation density, the incompressibility of symmetric matter has been determined from giant monopole resonances of heavy nuclei and found to be $K_{nm}(\rho_0) = 230 \pm 10$ MeV [14]. Various EOS calculated for symmetric nuclear are shown in Figure 4 as function of density [15].

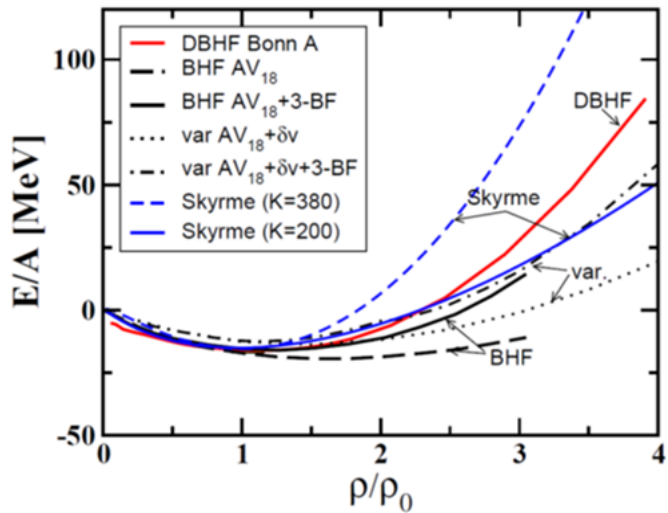

**Figure 4.** The isospin-symmetric nuclear matter equation-of-state (EOS) from soft and hard Skyrme forces compared to the predictions from microscopic ab initio calculations [15].

High-energy heavy-ion collisions offer the possibility to study the nuclear EOS at densities beyond $\rho_0$. Pioneering experiments on the high-density EOS of symmetric nuclear matter have been performed at the Bevalac in Berkeley and at the AGS in Brookhaven, which measured the collective flow of protons in Au + Au collisions at kinetic energies from 0.4 to 11 A GeV [16,17,18]. The collective flow of hadrons is driven by the pressure gradient inside the reaction volume, and, hence, exhibits a sensitivity to the EOS. The results of these early experiments have been interpreted by BUU transport model calculations by tuning the value of the nuclear incompressibility [19]. However, the result is not conclusive: while the directed flow data can be explained assuming a soft EOS ($K_{nm} = 210$ MeV), the data on the elliptic flow point towards a stiff EOS ($K_{nm} = 300$ MeV). Only very soft or extremely hard equations-of-state have been ruled out by this analysis.

The EOS of symmetric matter has been also studied in heavy-ion collision experiments at the SIS18 accelerator at GSI, where densities up to about 2 $\rho_0$ are reached. The FOPI collaboration (**Four Pi**) investigated the elliptic flow of protons and light fragments in Au + Au collisions at beam energies up to 1.5 A GeV. The FOPI results are not compatible with the elliptic flow value measured at the Bevalac in Au + Au collisions at 1.15 A GeV but rather indicate a soft EOS. According to IQMD transport calculations with momentum-dependent interactions, the explanation of the FOPI data requires a nuclear incompressibility of $K_{nm} = 190 \pm 30$ MeV [20].

Another observable sensitive to the EOS is subthreshold particle production. For example, the production of strange particle in proton-proton collisions via the reaction p + p → $K^+ \Lambda$ p requires a minimum beam energy of 1.6 GeV. In heavy-ion collisions, however, $K^+$ mesons have been measured by the KaoS collaboration already at beam energies of 0.8 A GeV [21]. In such collisions, the associated production of kaons and lambdas proceeds via multiple-step interactions with Delta resonances and pions involved. Sequential binary collisions or even three-body collisions are favored at high densities, and, therefore, the kaon yield in heavy-ion collisions at subthreshold beam energies is enhanced for lower values of the nuclear incompressibility. The KaoS collaboration measured the yield of $K^+$ mesons in Au + Au and C + C collisions at beam energies between 0.8 and 1.5 A GeV. According to transport calculations, the $K^+$ yield in such a small system like C + C does not depend on the EOS, and, hence, can be regarded as a reference measurement, which can be used to reduce the systematic uncertainties both in experiment and model calculation. As a result, RQMD and IQMD transport model calculations could reproduce the KaoS data when assuming a soft EOS with a nuclear incompressibility of about 200 MeV and taking into account momentum-dependent interactions [22,23].

The FOPI and the ASY-EOS collaborations have also measured the elliptic flow of neutrons and protons in heavy-ion collisions at GSI [24,25] in order to study the symmetry energy $E_{sym}(\rho)$. The experimental data of the FOPI collaboration have been compared to the results of UrQMD transport code calculations, which extracted a value for the symmetry energy of about $E_{sym} = 60 \pm 10$ MeV at 2 $\rho_0$. About twenty years later, the uncertainty of the FOPI result could be reduced by the ASY-EOS collaboration, which used an improved setup and found a value of $E_{sym} = 55 \pm 5$ MeV at twice saturation density. These experiments represent a substantial progress towards the determination of the symmetry energy at high densities, as so far the experiments measured $E_{sym}$ and L only at saturation density and these values then were used for extrapolations of $E_{sym}$ to higher densities [26].

In order to study the high-density EOS, which is relevant for the understanding of the most massive neutron stars, densities well above twice saturation densities should be produced and investigated in the laboratory. This is illustrated in Figure 6, which depicts the maximum mass of neutron stars as function of their central density for a variety of EOS [27]. The most massive neutron star discovered so far has a mass of $2.14 + 0.20 - 0.18$ solar masses [28], and, according to Figure 5, a central density between 4 and 8 $\rho_0$ depending on the EOS. Such densities can be reached in heavy-ion collisions for beam kinetic energies which will be available at the FAIR SIS100 synchrotron. According to model calculations, the nuclear fireball will be compressed to densities of about 5 $\rho_0$ in central Au + Au collisions already at kinetic beam energies of 5 A GeV and densities above 8 $\rho_0$ are reached at 10 A GeV [5].

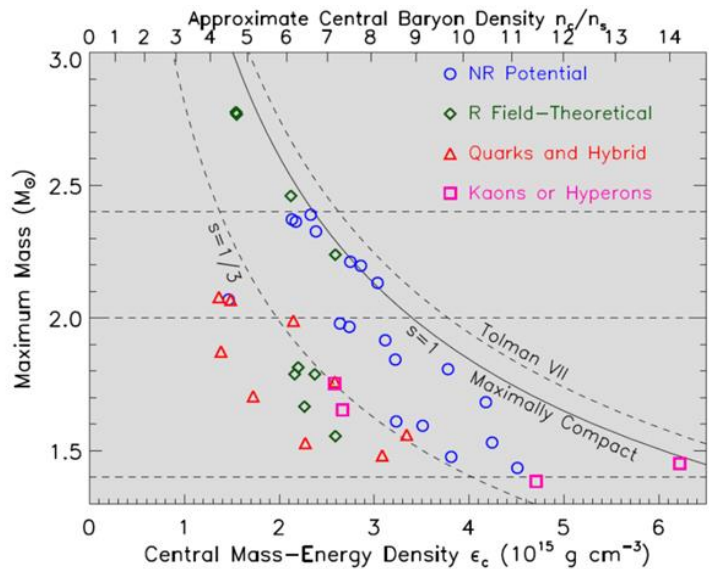

**Figure 5.** Maximum mass of neutron stars versus central density for different EOS [27].

An important part of the research program of the future Compressed Baryonic Matter experiment at FAIR is devoted to the exploration of the high-density EOS. This includes the measurement of the collective flow of protons, hyperons and light fragments at beam energies from 2 to 11 A GeV. Moreover, similar to kaon production at GSI-SIS18 energies, the subthreshold production of multi-strange (anti-)hyperons is expected to be sensitive to the high-density EOS at SIS100 energies. Transport models have investigated the production of Ξ and Ω hyperons at subthreshold beam energies and found, that the main source of multi-strange hyperons are strangeness exchange reactions involving kaons and lambdas [29,30]. These sequential collisions happen more frequently at high densities, and, therefore, the yield of multi-strange hyperons reflect the density in the fireball. The sensitivity of multi-strange hyperon production on the density of the reaction volume, and, hence, on the EOS, has been calculated with the novel PHQMD transport mode [31]. Preliminary results indicate, that in Au-Au collisions at FAIR energies the expected yields of $\Xi^-$ and $\Omega^-$ hyperons and even more of $\Xi^+$ and $\Omega^+$ anti-hyperons, are substantially higher for a soft EOS than for a stiff EOS. Therefore, the research program of the CBM experiment at FAIR includes the systematic study of multi-strange (anti-) hyperon production for different beam energies and beam-target combinations, in order to explore the high-density nuclear matter equation-of-state at neutron star core densities.

The understanding of neutron star matter requires, however, in addition to the knowledge of the high-density EOS for symmetric matter, also information about the symmetry energy $E_{sym}$ at high densities. As mentioned above, one option is to measure the elliptic flow of neutrons and charged particles at FAIR energies. Moreover, particles with opposite isospin, that is, with $I_3 = \pm 1$, produced in heavy-ion collisions, are regarded to be sensitive to the neutron and proton distributions in the reaction volume. In particular, attempts have been made to extract information on the $E_{sym}$ from the $\pi^-/\pi^+$ ratio measured in Au + Au collisions at 400 A MeV [32]. However, it turned out, that the sensitivity of this observable decreases strongly for beam energies above the pion production threshold. In addition, the $\pi^-/\pi^+$ ratio also depends on the unknown $\Delta(1232)$ in-medium potentials. In heavy-ion collisions at beam energies between 2A and 11A GeV, the $\Sigma^-/\Sigma^+$ ratio may by a promising observables sensitive to $E_{sym}$ at high baryon densities. In this case, the experimental challenge is to reconstruct the $\Sigma$ hyperons, which decay in a charged and a neutral particle, by the missing mass method based on the precise measurement of the tracks of the $\Sigma$ and its charged daughter.

*3.2. Searching for New Phases of QCD Matter at High Net-Baryon Densities*

The conjectured QCD phase diagram shown in Figure 3 indicates a first order phase transition at high net-baryon densities, ending in a critical point. None of these structures has been found so far. As mentioned in the beginning, first-principle theories, such as perturbative QCD, still fail to make reliable predictions for properties of matter at large baryon chemical potentials. In order to get an idea on the degrees-of-freedoms of matter in the core of neutron stars, we have to rely on effective models. One example is illustrated in Figure 6, which depicts a schematic picture of a transition from nuclear to deconfined quark matter with increasing density [33]. For densities below 2 $\rho_0$, the dominant interactions occur via a few meson or quark exchanges and the matter can be described in terms of interacting nucleons. For densities in the range from 2 $\rho_0$ to 4 – 7 $\rho_0$, many-quark exchanges dominate, and the system gradually changes from hadronic to quark matter. For densities beyond 4 – 7 $\rho_0$, nucleons percolate, start to melt and to dissolve into their constituents, which do not longer belong to particular nucleons. A perturbative QCD description is expected to be valid only for extremely high densities, that is, above 100 $\rho_0$.

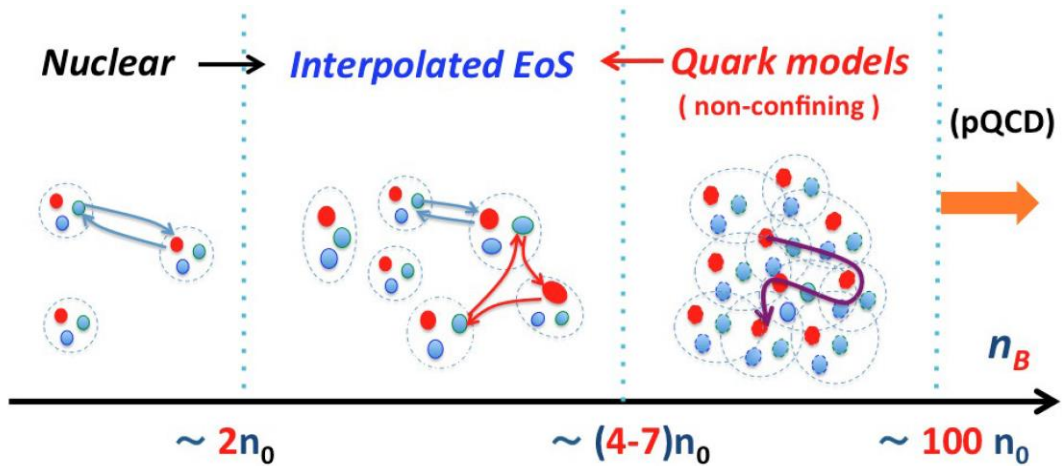

**Figure 6.** Sketch of a suggested continuous the transition from nuclear to quark matter with increasing density in units of saturation density (see text). Figure courtesy of T. Kojo [33].

Various calculations have been performed which indicate phase transitions from hadronic to quark matter in the core of massive neutron stars, in neutron star mergers or even in core collapse supernovae of very massive stars. Calculations based on a non-local 3-flavor Nambu Jona-Lasinio model predict a mixed phase of hadrons and quarks in the core of a 2 solar mass neutron star for densities of about 5 $\rho_0$ and even a transition to pure quark matter above 8 $\rho_0$ [34]. According Chiral Mean Field model simulations of a neutron star merger with a total mass of 2.8 solar masses, rather high densities and temperatures are expected to be reached [35]. The calculation predicts a phase transition to pure quark matter at a density of 4 $\rho_0$ and at a temperature of about 50 MeV. This phase transition occurs shortly before the high-mass neutron star collapses into a black hole. In order to explain the core-collapse supernovae of a blue-supergiant with 50 solar masses, calculations have to take into account a first order phase transition from nuclear matter to the quark-gluon plasma at densities between 3 and 4 $\rho_0$ for symmetric matter. The remnant of such a stellar explosion is a neutron star of about 2 solar masses with a quark matter core [36].

The question, whether there is (i) a smooth transition from baryon to quark matter ("quarkyonic matter") at densities above 4 – 7 $\rho_0$ as suggested by Reference [33] or (ii) a mixed phase of hadrons and quarks between 4 and 8 $\rho_0$ [34] or (iii) already pure quark matter is created above 4 $\rho_0$ as predicted the simulations of neutron star mergers [35] or (iv) a first order phase transition occurring in symmetric matter at densities between 3 and 4 $\rho_0$ [36] or (v) still hadronic matter exists at this density, might be decided by heavy-ion collision experiments. As will be discussed in the following, the systematic measurement of hadronic observables and lepton pairs in heavy-ion collisions is a very promising strategy in order to obtain a consistent picture of the properties of QCD matter at high net-baryon densities.

*3.3. Probing the Fireball Temperature with Di-Leptons*

The temperature of the fireball produced in heavy-ion collisions can be determined from the invariant mass spectra of di-leptons. This has been recently demonstrated by an experiment at GSI-SIS18 performed by the HADES collaboration, which studied the production of electron-positron pairs in Au + Au collisions at a beam energy of 1.25 A GeV. In such reactions, it is predicted that baryon densities of up to about 2.5 $\rho_0$ are created. The di-electron invariant mass spectrum has been corrected for the combinatorial background and for the contributions from known vector meson decays. The result of the subtraction procedures is the so called access yield, which can be described by a thermal distribution with a temperature of about 72 MeV, and, hence, provides the possibility to directly determine the temperature of the fireball [37]. Therefore, the precise measurement of the dilepton access yield, as demonstrated by the HADES collaboration, opens the unique possibility to determine the caloric curve, when measuring the excitation function of the fireball temperature. This would be the first direct experimental signature for a phase transition in high-density nuclear matter. It is worthwhile to note, that at FAIR energies also the invariant mass range above 1 GeV/c$^2$ becomes accessible, where the dilepton yields from vector meson decays do not contribute [38]. This is illustrated in the left panel of Figure 7, which depicts a simulation of various sources of dilepton radiation, including a contribution from the quark-gluon plasma, for central Au + Au collisions at 20 A GeV. The right panel of Figure 7 sketches the fireball temperature as function of collision energy for two scenarios. The (red-dashed) curve illustrates the temperature extracted from the dilepton invariant mass range between 1–2 GeV/c$^2$, as calculated with a fireball model and a coarse-graining approach [39]. Such a smooth temperature increase is expected for hadronic matter or a crossover transition. The purple line illustrates a hypothetic caloric curve reflecting a first order phase transition. In addition, results from the HADES [37] and NA60 [40] experiments are shown. The NA60 data point is also extracted from the spectral slope above an invariant mass of 1 GeV/c$^2$.

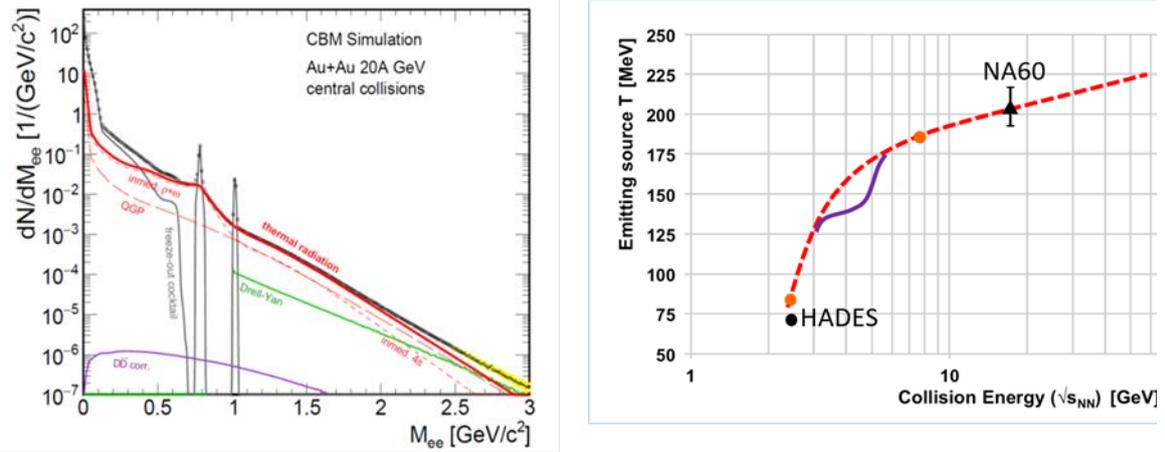

**Figure 7.** Left panel: Di-electron invariant mass spectrum simulated for central Au + Au collisions at 20A GeV. Right panel: Sketch of the emitting source temperature as function of collision energy for two scenarios. The red dashed curve is the result of a calculation based on a thermal model assuming no phase transition or a crossover (see text) [39]. The purple line illustrates a hypothetical caloric curve reflecting a first order phase transition. In addition, the experiments result from HADES [37] (**H**igh-**A**cceptance **D**i-**E**lektron **S**pectrometer) and NA60 [40] are shown.

The information provided by the excitation function of the fireball temperature should be complemented with results from measurements of hadronic observables, which are also expected to shed light on effects associated with a phase transition, such as the onset of deconfinement and the critical endpoint. The goal must be to measure a set of diagnostic probes, which finally provide conclusive evidence for a consistent scenario. Promising candidates for such hadronic observables are discussed in the following.

*3.4. Searching the Onset of Deconfinement with Multi-Strange Hyperons*

The statistical hadronization model describes well the measured yields of hadrons and even of light (anti-) nuclei produced in ultra-relativistic heavy-ion collisions, supporting the scenario of a chemically and thermally equilibrated fireball, which freezes-out at a temperature of about 156 MeV [41]. The fact, that also the yields of $\Omega^-$ and $\Omega^+$ hyperons agree with the model assumptions, although the hyperon-nucleon scattering cross section is small, was interpreted as a signature for a phase transition from the quark-gluon plasma to a hadronic final state, which drives multi-strange hyperons into equilibration [42]. The yields of multi-strange hyperons measured in Pb + Pb collisions at a beam kinetic energy of 30 A GeV at the CERN-SPS were also found to be in agreement with the statistical model, which in this case extracts a freeze-out temperature of 138 MeV [43]. Even for a smaller collision system such as Ar+KCl and a much lower beam energy of 1.76 A GeV, the abundantly produced hadrons agree with the thermal model for a temperature of about $T = 76$ MeV [44]. However, the yield of $\Xi^-$ hyperons was found to exceed the statistical model by about a factor $24 \pm 9$, indicating a non-equilibrium production mechanism of $\Xi^-$ hyperons. The CBM experiment will carefully investigate the yield of multi-strange hyperons produced in A+A collision at SIS100 energies and search for the lowest beam energy where all produced hadrons including $\Xi$ and $\Omega$ hyperons are equilibrated, which might indicate the onset of deconfinement in QCD matter at high net-baryon densities.

*3.5. The Quest for A First Order Phase Transition with Fluctuations of Conserved Quantities*

Model calculations predict, that in heavy-ion collisions an effect like critical opalescence should occur, if the particles freeze-out in the vicinity of the critical point [45]. Such a phenomenon could be observed as event-by-event fluctuations of the multiplicity distributions of conserved quantities such as baryon number, strangeness and electrical charge, measured in a limited rapidity range. In particular, the higher moments of these distributions are expected to exhibit an increased sensitivity to critical phenomena. The STAR collaboration has performed a beam energy scan of various observables, including net-proton distributions measured event-by-event and found an increase of the 4[th] order cumulant (kurtosis) towards the lowest collision energy of $\sqrt{s_{NN}}$ = 7.7 GeV [46]. However, in order to discover the maximum of the fluctuation, and, hence, the possible location of the QCD critical endpoint, the measurement has to be extended towards lower collision energies. This is hardly possible at RHIC because of the steeply decreasing luminosity of the collider but is an important part of the research program of the CBM fixed-target experiment at FAIR. In case, the maximum of the 4[th] order cumulant of the net-proton multiplicity distribution will be found at top FAIR energies, then this observation should be corroborated by dilepton measurements evidencing a caloric curve as indicated in the right panel of Figure 9, before claiming discovery of the QCD critical endpoint.

*3.6. Hyperons in Dense Nuclear Matter*

The high-density EOS might not only be softened by a phase transition from hadronic to quark degrees of freedom but also by hyperons, which should appear at high densities, if the chemical potential of neutrons and protons exceeds the chemical potential of hyperons. This process would prevent the formation of massive neutron stars, which, however, have been observed. In order to solve this "hyperon puzzle," various models propose different mechanisms to circumvent a softening of the EOS at high densities. For example, calculations based on a 3-flavor Nambu Jona-Lasinio model describe quark deconfinement in neutron star cores and predict a mixed phase, where hadrons including hyperons and quarks coexist [34]. In this case, repulsive vector interactions among the quarks are proposed to provide a stiff EOS. The results of calculations based on chiral effective field theory assuming only hadronic degrees-of-freedom are illustrated in Figure 8, which depicts the chemical potentials of neutrons and lambdas as function of density [47]. The model features a repulsive ΛN potential together with repulsive 3-body ΛNN interactions at high densities, and, thus, prevents the condensation of lambdas below densities of about 5 $\rho_0$.

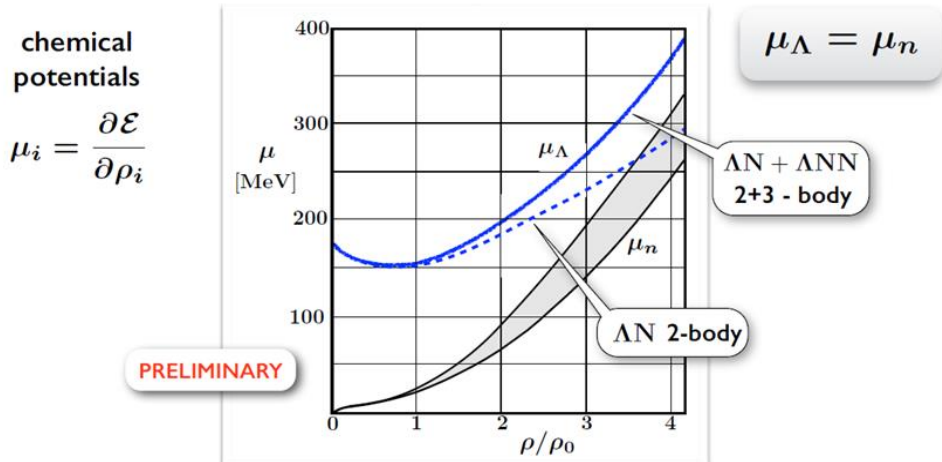

**Figure 8.** Chemical potentials of Λ hyperon $\mu_\Lambda$ and neutron $\mu_n$ in neutron matter as a function of baryon density. The lambda chemical potential is calculated taking into account ΛN two-body interactions only (blue-dashed line) and including ΛNN three-body forces (blue solid line). The grey-shaded area between the black lines indicates the neutron chemical potential [47].

The research program of CBM includes the study of ΛN, ΛNN and ΛΛ by the measurement of hypernuclei produced in heavy-ion collisions. According to calculations based on a UrQMD-hydro hybrid model and on a hadronic cascade, the yield of light (double-) lambda hypernuclei exhibits a maximum in the FAIR energy range [48]. This maximum is due to the fact, that the hyperon yield increases with increasing beam energy, while the yield of light nuclei decreases with increasing beam energy. Because of the extremely high reaction rate, the CBM experiment is well suited for the discovery of yet unknown (double-) lambda hypernuclei and for the determination of their lifetime.

## 4. The Compressed Baryonic Matter Experiment at FAIR

The mission of the Compressed Baryonic Matter (CBM) experiment is to explore the properties of QCD matter at neutron star core densities. Such densities can be produced for a time span of about 5–10 fm/c in collisions between two heavy nuclei at FAIR energies. As discussed above, during this very short period of time, new particles are created and emitted, which serve as messengers from the sense fireball. The experimental challenge is to identify these partly very rarely produced diagnostic probes and to measure their yield and momentum with high precision and unprecedented statistics. In order to fulfil this requirement, a fixed-target setup with fast and radiation hard detectors coupled to a high-speed data read-out and acquisition system has been developed, which is able to run at reaction rates of up to 10 MHz. It is worthwhile to note, that existing and heavy-ion experiments under construction are operated at rates up to 10 kHz at colliders and up to 50 kHz in fixed target mode [49].

At FAIR, heavy-ion collisions will be investigated by the HADES experiment and the CBM experiment. The two setups will be operated alternatively and are shown in Figure 9. The HADES detector system features a large polar angular acceptance between 18 and 85 degrees, and, hence, is very well suited for reference measurements with proton and ion beams of intermediate masses like Ag at low SIS100 energies. The HADES setup will measure hadrons and electron-positron pairs. During HADES operation, the beam will be stopped by a beam dump located in front of the CBM magnet, as sketched in Figure 9. For measurements with the CBM setup, this beam dump will be removed.

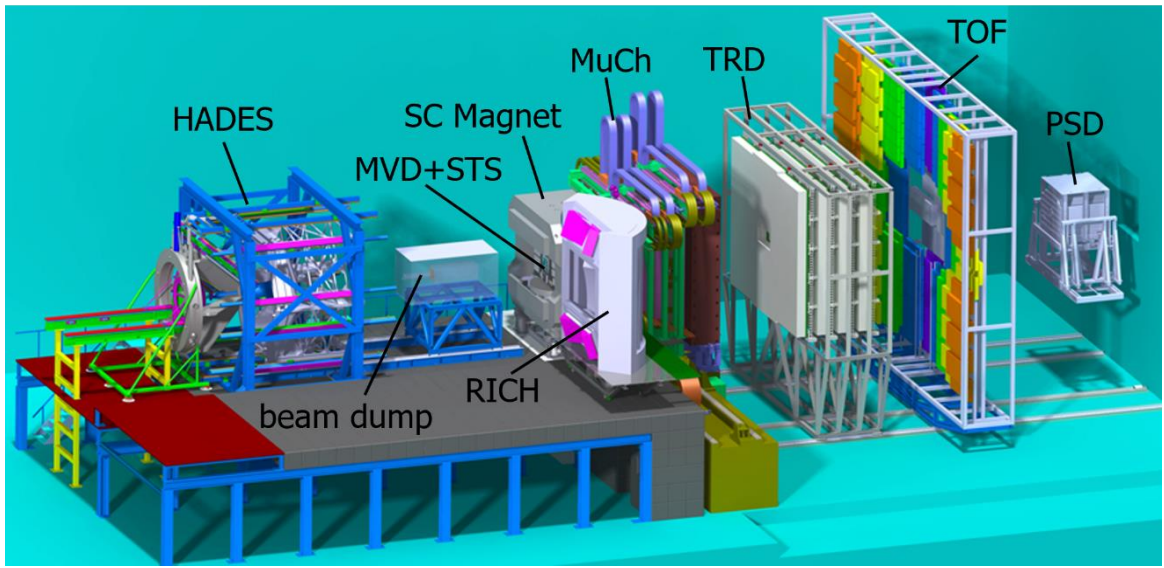

**Figure 9.** The HADES detector (left) with its beam dump, which will be removed during Compressed Baryonic Matter (CBM) operation. The CBM experimental setup comprises the following components: the superconducting dipole magnet, the Micro-Vertex Detector (MVD) and Silicon Tracking System (STS) located in the magnet gap, the Muon chamber (MuCh) system in measuring position, the Ring Imaging Cherenkov (RICH) detector in parking position, the Transition Radiation Detector (TRD), the Time-Of-Flight (TOF) detector and the Projectile Spectator (PSD) detector (see text).

The CBM is designed to identify hadrons, electrons and muons in Au+Au collisions at beam energies from 2 to 11A GeV at reaction rates up to 10 MHz. Under such conditions, up to about $10^9$ charged particles have to be measured per second. This requires fast and radiation hard detectors and a novel data read-out and acquisition system. The CBM detector system accepts polar angles from 2.5 to 25 degree, which is sufficient to cover the forward rapidity range including mid-rapidity. The momentum of charged particles will be determined by measuring their tracks with a Silicon Tracking System (STS) comprising eight stations located in the field of a large-aperture superconducting dipole magnet, at a distance between 30 and 100 cm downstream the target. The STS is composed of about 900 double-sided micro-strip silicon sensors with a pitch of 58 μm, a width of 60 mm and a length between 20 and 120 mm, depending on the track density. In total, about 1.8 million channels will be read out a free-streaming, trigger-less front-end electronics, based on an ASIC, which provides a time-stamp to each detector signal. In order to identify the decay vertex of very short-lived particles such as charmed mesons, which decay within a few hundred μm behind the target, a Micro-Vertex Detector (MVD) is installed between the target and the first STS station, comprising four layers composed of Monolithic Active Pixel Sensors (MAPS).

In order to determine the mass of the tracked particles, a Time-Of-Flight (TOF) wall is installed about 7 m downstream the target, with an active are of about 100 m$^2$. The TOF wall consists of Multi-gap Resistive Plate Chambers (MRPCs). The MWPC modules located at small polar angles, where the particle density is high, are built from low-resistivity glass and can be operated at rates up to 20 kHz/cm$^2$. The overall time resolution of the TOF detector is 80 ps. A Ring Imaging Cherenkov (RICH) detector will be installed after the magnet for the identification of electrons and positrons up to momenta of about 8 GeV/c. In order to separate electrons from pions at higher momenta, a Transition Radiation Detector (TRD) will be installed downstream the RICH. The pion suppression factor of RICH together with TRD is about $2 \cdot \times 10^4$ for momenta between 2 and 6 GeV/c and reduces to a value of about 80 for momenta above 10 GeV/c, where mostly the TRD is effective, with some contribution from the TOF wall.

For muon measurements, the RICH will be replaced by a Muon Chamber (MuCh) system. The MuCh consists of one hadron absorber made of graphite, followed by up to four iron plates, with muon tracking detector triplets behind each absorber. The circular detector triplets behind the first two absorbers, where the particle rate still is up to 500 kHz/cm$^2$, are composed of 18 and 20 trapezoidal Gas Electron Multiplier (GEM) detectors per layer. The third and fourth triplet, where the particle rate is decreased to about 15 and 4 kHz/cm$^2$, will be built from single-gap RPC detectors with low-resistivity Bakelite electrodes. Behind the fifth iron absorber, the muons will be tracked by the TRD and finally identified by the TOF. The event characterization, that is, the determination of the reaction plane angle and the centrality, will be performed by the Project Spectator Detector (PSD), a segmented hadron calorimeter located about 10 m downstream the target.

For the high-rate CBM experiment, the data read-out and acquisition system plays a crucial role. As mentioned above, the time-stamped signals will be read out without event correlation and transferred to a high-performance computing farm, the GSI GreenIT Cube, where online event reconstruction and selection is performed by high speed algorithms. In a first step, the tracks of the charged particles were reconstructed from the space and time information of the various detector signals. In a second step, the particles will be identified, taking into account secondary decay vertices and information of RICH or MuCh, TRD and TOF. Finally, the particles will be grouped to events, which will be selected for storage if they contain important observables. In parallel, the event is characterized using information from the PSD.

## 5. Summary

The research programs at the future Facility for Antiproton and Ion Research include the investigation of fundamental open issues of astrophysics, such as the nucleosynthesis in the universe and the properties of dense nuclear matter relevant for our understanding of neutron stars, neutron star mergers and supernovae. Experiments at the superconducting Fragment-Separator will study the origin of the heaviest elements by measuring the properties of very short-lived neutron-rich or neutron- deficient isotopes, which play a decisive role in the rapid neutron capture and rapid proton capture processes in stellar explosions or collisions. The CBM experiment is designed to explore the QCD phase diagram in the region of high baryon densities, including nuclear matter equation-of-state and possible phase transition from hadronic to quark matter. These laboratory measurements will complement astronomical observations devoted to the determination of the mass and radius of neutron stars, which also will constrain the EOS at neutron star core densities. In addition, the heavy-ion experiment will shed light on the degrees-of-freedom of high-density QCD matter, which is hardly accessible from astrophysical measurements. Finally, the investigation of (double-) lambda hyper-nuclei, which are predicted to be abundantly produced in heavy-ion collisions, will allow to study ΛN, ΛNN and ΛΛ interactions, and, hence, will contribute to the solution of the "hyperon puzzle" in neutron stars.

In order to execute the research program outlined above, high-precision multi-differential measurements of diagnostic probes including strange particles, lepton pairs and hyper-nuclei have to be performed. This requires to operate the detectors and the data-acquisition system at extremely high reaction rates of up to 10 MHz. To this end, substantial technological efforts have been undertaken during the last decade, including the development of fast and radiation hard detectors, of a trigger-less, free streaming read-out electronics, and, last but not least, of high-speed algorithms tailored to run on modern computer architectures, providing online event reconstruction and selection.

**Acknowledgments:** The development of the CBM experiment is performed by the CBM Collaboration, which consists of more than 470 persons from 55 institutions and 12 countries. The CBM project is supported by the German Ministry of Education and Research, the Helmholtz Association and national funds of the CBM member institutions. The project receives funding from the Europeans Union's Horizon 2020 research and innovation programme under grant agreement No. 871072. The author acknowledges support from the Ministry of Science and Higher Education of the Russian Federation, grant N 3.3380.2017/4.6 and by the National Research Nuclear University MEPhI in the framework of the Russian Academic Excellence Project (contract No. 02a03.21.0005, 27.08.2013).